# An Incomplete Cryptography based Digital Rights Management with DCFF

Ta Minh Thanh

Department of Computer Science
Tokyo Institute of Technology
2-12-2, Ookayama, Meguro,
Tokyo, 152-8552, Japan.
Email:thanhtm@ks.cs.titech.ac.jp

Munetoshi Iwakiri

Department of Computer Science
National Defense Academy
1-10-20, Hashirimizu, Yokosuka,
Kanagawa, 239-8686, Japan.
Email: iwak@nda.ac.jp

*Abstract*—In general, DRM (Digital Rights Management) system is responsible for the safe distribution of digital content, however, DRM system is achieved with individual function modules of cryptography, watermarking and so on. In this typical system flow, it has a problem that all original digital contents are temporarily disclosed with perfect condition via decryption process. In this paper, we propose the combination of the differential codes and fragile fingerprinting (DCFF) method based on incomplete cryptography that holds promise for a better compromise between practicality and security for emerging digital rights management applications. Experimental results with simulation confirmed that DCFF keeps compatibility with standard JPEG codec, and revealed that the proposed method is suitable for DRM in the network distribution system.

## I. INTRODUCTION

Recent technological advances in the area of multimedia content distribution have resulted in a major reorganization of this trade. Valuable digital artworks can be reproduced and distributed arbitrarily without any control by the copyright holders. Digital content has these advantages but has shortcoming such as piracy of copyright, an illegal copy and editing. Thus, issues related to intellectual property rights protection and management arise.

DRM system can prevent from thus piracy of the contents as cutting the user off from an illegal approach to the content through the encryption of it [1, 2]. However, conventional DRM technologies are manipulated by encryption and watermark method separately. Therefore, original content is disclosed temporarily inside user's system in the user's decryption process (key management process) [3]. In that case, the original content is able to be distributed without any fingerprinting information. And producer also can't trace the illegal distributor.

To ensure that digital content is used for its intended purpose after it has been delivered to customers often requires the ability to track and identify entities involved in unauthorized redistribution of multimedia content. Digital fingerprinting is a technology for enforcing digital rights policies whereby unique labels, known as digital fingerprints, are inserted into content prior to distribution. For multimedia content, fingerprints can be embedded using conventional watermarking techniques to an individual.

If fingerprinting system is unifiedly integrated with DRM system, however, it is possible to trace the illegal traitor and protect the content having the copyright. Fingerprinting system plays the role of the insertion of user information into the content which is similar to the watermarking technology [4, 5, 6].

In this paper, we describe an integration and implementation of DRM technique using DCFF based on an incomplete cryptography system. The know-how of the proposed method is the fundamental incomplete cryptography. Using incomplete cryptography, the producer will encode a part of original contents to make scrambled contents (trial contents) for distributing widely to users via network. On the other hand, in the receivers/users side, the quality of scrambled contents can be controlled with the fingerprinted key at the incomplete decoding process for embedding individual user information into the incomplete decoded contents simultaneously.

This paper is organized as follows. Section 2 introduces some reviews of related works for DRM applications. The proposed incomplete cryptography is presented in Section 3. Section 4 explains the detail of algorithm for DCFF and implementation to DRM system. The application of DCFF in JPEG (Joint Photographic Experts Group) algorithm and those experimental results are described in Section 5. The conclusion is summarized in Section 6.

## II. REVIEW OF RELATED WORKS

Our research has been inspired by a number of conventional works available in the literature that employ digital watermarking for copyright protection of digital content. We focus on the dual targets of digital content: legal access and traitor tracing. Many papers have been written about the techniques that use both encryption and fingerprinting in the literature. The existing integration of encryption and fingerprinting technology, divided into three categories, is briefly described as follows.

Server-side encryption and fingerprint embedding: A original content is separately embedded with an user's fingerprint and then encrypted with a global key to create an encoded content. However, there are some disadvantages in this model: The first one is inefficient bandwidth utilization because digital fingerprint embedding is done at the server, then it will be high computation cost if user repeats request of same content many times. The second one is insecure because only a single global encryption key is used. If a malicious user can decode another user's data, then original content can be obtained and be illegally distributed.

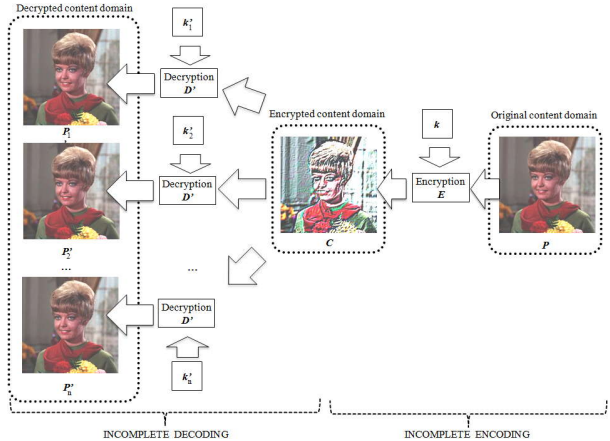

Fig. 1. Overview basic idea of incomplete cryptography.

Server-side encryption and user-side fingerprint embedding (conventional DRM system): Macq *et al.* [8] proposed the fingerprint embedding at the users side for digital TV. After that, Hartung *et al.* [9] and Bloom *et al.* [10] extended the idea of [8] for applying to DRM in digital cinema. In this model, only one global key encryption is employed to encode the digital content at the server side. Next, the encoded content can be sent to different users via network for multicasting. At the user side, the encoded content can be decrypted according to the global key and it must to be fingerprinted based on user's information. Meanwhile, the digital fingerprint must be embedded into the content after decoding to generate the fingerprinted content for each user. In general, a watermarked software (DRM controller software) is necessary in this scenario for embedding the fingerprinting information into content after decoding the encrypted content. However, the watermarked software is still an open problem because *the original content is possibly revealed inside the system by this software.*

Joint fingerprinting and decryption (JFD): Kundur *et al.* [7] proposed a JFD method to reduce system complexity and achieve the real-time requirement. In JFD method, the encoded content is partially decrypted such that the un-decrypted parts imitate multimedia fingerprint embedding. JFD is conceptually promising, achieving partial multimedia decryption and multimedia fingerprint embedding at the same time. However, the un-decrypted parts should not affect the whole transparency of the fingerprinted content and it should be robust against some attacks.

In next section, we present a new incomplete cryptography method, which can be incorporated into the DRM system. Most importantly, our method does not encounter the same problems as the above three types of methods.

### III. PROPOSAL OF INCOMPLETE CRYPTOGRAPHY

The proposed incomplete cryptography for DRM systems is explained in this section. There are two steps in the proposed cryptography: the incomplete encoding and the incomplete decoding. The basic idea of the incomplete cryptography is shown in **Fig. 1**.

#### A. Incomplete encoding

The incomplete encoding process is presented in this sub-section. Producer $T$ has a digital content $P$ and needs to create an encoded content by incomplete cryptography. In that case, $P$ will be encoded based on the encoder function $E$ with the encoder key $k$ to make the scrambled content $C$.

$$C = E(k, P) \qquad (1)$$

In the incomplete cryptography, $C$ can be simply recognized a part of $P$ (even if $C$ is not decoded). This feature is called *incomplete confidentiality*. $T$ can distribute $C$ widely to users as trial content via network.

#### B. Incomplete decoding

The incomplete decoding process is different from the complete decoding process. Decoded content is created by another decryption function $D'$ with another decoded key $k'_i$ ($i = 1,2,...,n$). Note that, $D'$ with $k'_i$ is different from $D$ with $k$. Decoded content $P'_i$ is calculated as follows:

$$P'_i = D'(k'_i, C) \qquad (2)$$

In this case, because $P'_i$ is decoded by another decryption function $D'$ with key $k'_i$, it will be different from original content $P$. Therefore, the relationship of $P$ and $P'_i$ is $P'_i \neq P$ in incomplete cryptography system. Hence, this decoder process is quite different from the complete cryptography. This feature is called *incomplete decode*.

According to the features of incomplete cryptography, if a set of decoder key $k'_i$ with decoder function $D'$ to decode the scrambled content $C$ are chosen, a set of decoder content $P'_i$ will be created and different from each other. Thus, if the incomplete cryptography is implemented to construct a distribution system via network, the producer can distinguish the legal user by $P'_i$ that is decoded based on key $k'_i$.

## C. New DRM system using Incomplete Cryptography

Since the the incomplete cryptography has the incomplete confidentiality and the incomplete decode feature, it makes no value to a secret transmission system. However, the incomplete cryptography is able to manipulate the contents via an information transmission system so far as the distortion level permits. Especially, if the features of the incomplete cryptography are implemented to a DRM system, the incomplete encoding (make scrambled content) and the incomplete decoding (make fingerprinted content) processes are required to resolve the above mentioned problem.

TABLE I.          COMPARISON OF EXIST METHODS.

|  | Watermark | Fingerprint | JFD [7] | Ours |
|---|---|---|---|---|
| Embedding | Watermark | Hash | Un-decrypted | User ID |
| Trial content | × | × | ○ | ○ |
| Cost | × | × | ○ | ○ |
| Identification | Watermark | Hash | Un-decrypted | User ID |
| Traceability | × | △ | △ | ○ |

Note: "×" means Bad; "△" means Good; "○" means Very Good

The idea of new DRM system based on incomplete cryptography is presented in this subsection [11, 12]. A DRM system requires to enable the distribution of original contents safely and smoothly, as well as to enable the secondary use of contents under rightful consents. When a DRM system is constructed using the incomplete cryptography to implement a content distribution system, it is not only the safety distribution method to users, but also the solution of the conventional DRM problem (disclose the original content inside user's system).

Before distribution, $T$ has a digital content $P$ and needs to be sent to users as much as possible. Thus, $T$ creates a scrambled content $C$ with the encryption key $k$ based on the incomplete encoding. $C$ is to disclose a part of $P$. It means that $C$ is maintained over the minimum quality of $P$. $T$ distributes $C$ to users widely via network as a trial content.

After trial $C$ which is distributed via network, $R$ decides to purchase a digital content. Then, $R$ has to register his/her individual information (user ID). This information will be used as the fingerprinting information ($w_m$) and embedded into the content. When $T$ receives the purchaser's agreement, $T$ sends a fingerprinted key $k'_i$ to $R$. $k'_i$ is the incomplete decoding key and it is prepared individually to each user.

$R$ decodes $C$ using $k'_i$ and obtains the high quality content $P'_i$. In this decoding process, ID information ($w_m$) of user will be embedded in $P'_i$ as the copyright information.

Therefore, when a producer wishes to check whether the user is a legal user, he/she can extract the fingerprinting information from $P'_i$ and compare with his user database. If the fingerprinting information matches his database, the user is a legal user. Conversely, if the fingerprinting information is a different from his database, the user is an illegal user. Furthermore, it can specify to trace the source of pirated copies. The purpose of this proposed method is to inform the users about the existence of fingerprinting which can exactly identify users, and limit the illegal redistribution in advance.

*Note that*, in this study, robustness is not the major concern. Therefore, we derive analytic bounds of the embedded signals to achieve the highest transparency and ensure that our technique can trace the traitor exactly.

## D. Comparison of our method and conventional method

There are three primary technologies currently used to identify content in the new unstructured distribution: *digital watermarking*, *digital fingerprinting* and *JFD*. While three enable content identification, they differ in some significant ways that bear on their appropriateness for different applications. Our proposed method differs from those techniques. **Table I** shows the brief comparison of conventional methods with our proposed method.

We compare our method with conventional methods on a number of criteria to provide a better understanding of the advantages of proposed approach.

- Embedding information: is defined here as the information is embedded into digital content after distribution to users. *Watermarking* is generally using copyrights information for embedding. Watermarking is not intended to identify the user. The presence of the watermark is to prove that the content is a copy. *Fingerprinting* is generated by content-based processing such as fingerprint of human. Normally, the hash value of digital content is used to distinguish the contents. In *JFD* method, the un-decrypted parts imitate multimedia fingerprint embedding in the decoding process. On the other hand, in *our method*, the user ID is used to embed into decoded content for individual user when user decodes the trial content.

- *Trial content*, is the low quality contents, which are distributed to users widely via network. In conventional *watermarking* and *fingerprinting* method, there is not the concept of the trial content. The copyrights information is directly embedded into the content before delivering to the user. Otherwise, *our method* and *JFD* method employ the trial content for advertisement of content. Then, users can try the content beforehand to decide whether purchase or not.

- *Cost*, is considered the objective costs like computer resources and more subjective factors such as system complexity. In *watermarking* and *fingerprinting* method, since embedding information is directly embedded into the original content before distributing, server-side is almost responsible for embedding by high computation cost. If watermarking and fingerprinting is used for distinguishing the legal users, the user-based information should be embedded into content before distribution. Therefore, the cost computation of these methods are not good. On the other hand, *our method* and *JFD* shift the decoding process to user-side, then high computation cost in server-side is clearly reduced.

- *Illegal user identification*, is the measure method to detect whether the users is an illegal user. In *watermarking* method, if producer can detect the watermarking information from redistributed content, the user who possessed that content will be judged as the illegal user. In *fingerprinting* and *JFD* method, when detected fingerprint doesn't match against a reference database or un-decrypted parts doesn't match against database, the content possessor will be judged as illegal user. Otherwise, in *our proposed method*, we define that when user ID is not detected or doesn't match against database, then user is judged as illegal user. Therefore, **all attacks that try to remove or replace the user ID from an embedded content will be disabled. And the content, which is generated by those attacks will be considered as illegal content**.

- *Traceability*, is the ability of system for detection the source of pirated content. *Watermarking* method is generally used for copyrights protection, then watermark information is not particular user information. So, watermarking cannot trace the source of pirated content. *JFD* method is considered that it can trace the source of pirated content by detection the un-decrypted parts of individual user in decoded content, but it seems quite complicated. Unlike watermarking and JFD method, user ID is employed for information embedding in *our proposed* method and *fingerprinting*. So that, if we detect the user ID from decoded content, we can trace exactly the illegal source of pirated content.

## IV.  ALGORITHM OF DCFF

To demonstrate the practical compromises necessary for new DRM design, we propose a straightforward DCFF algorithm. In addition, we introduce the hash value [13] of the incomplete decoded content to distinguish the users. There are two basic processes in the incomplete cryptography with respect to digital content as shown in **Fig. 2** (a) and (b).

### A.  Scrambled method using differential codes

In this scrambled method, the incomplete encoding is utilized to obtain the scrambled content (trial contents) that is delivered to users via network. The scrambled content has an important role in user's decision of purchase. Suppose $P$ is a content in the original contents domain, $k = \{k^\xi_C, k^\psi_C\}$ is an encoder key. $C$ is archived by degrading the quality of $P$ (see Fig. 2 (a)). The encoding process is summarized as follows.

**Step 1**. Obtain $n$ significant coefficients from $P$, which are shown as:

$$P = \{p_0, p_1, \ldots, p_n\} \tag{3}$$

**Step 2**. To scramble the $n^{\text{th}}$ coefficient of $P$, a value of $k^\xi_C$ is randomly generated by $T$.

**Step 3**. The element $p_n$ is encrypted by $k^\xi_C$ with the operator $\odot$ to obtain the encoded coefficient is termed as scrambled coefficient $e_n$.

$$e_n = p_n \odot k^\xi_C \tag{4}$$

Reader should note that we require ($n$ - 1) operators $\odot$ that are randomly generated by key $k^\psi_C = \{-1, 1\}$. Operator $\odot$ is selected the subtraction (-) if $k^\psi_C = -1$ or the addition (+) if $k^\psi_C = 1$.

**Step 4**. Excepting the coefficient $p_n$, another coefficients $p_i$ ($0 \le i < n$) are encrypted to obtain another encrypted coefficient $e_i$ based on using $p_i$ with the previous difference $e_{i+1}$ and it is following:

$$e_i = p_i \odot e_{i+1}, \; i = 0,1,2,\ldots,n\text{-}1 \tag{5}$$

According to this algorithm, the significant coefficients in $P$ are encoded with an encoder key $k = \{k^\xi_C, k^\psi_C\}$ based on the differential codes algorithm and the scrambled content $C$ is obtained.

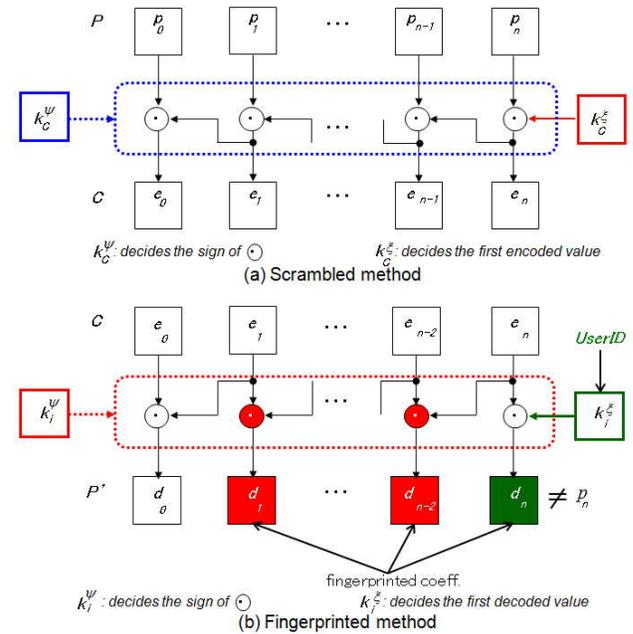

Fig. 2.  Overview of DCFF algorithm.

### B.  Fragile fingerprinted method

The fingerprinted method is utilized at the user $R_i$ side. Suppose $R_i$ receives the decoder key $k_i = \{k^\xi_i, k^\psi_i\}$ from $T$. The fingerprinted content is created by $R_i$ using the incomplete decoding process (Fig. 2(b)). This algorithm is following.

**Step 1**. Obtain $n$ significant coefficients from scrambled content $C$,

$$C = \{e_0, e_1, \ldots, e_n\} \tag{6}$$

**Step 2**. A user $R_i$'s information based decoder key $k^\xi_i$ that is created by $T$, is delivered to $R_i$.

**Step 3**. Then an element $e_n$ is decrypted by $k_i^\xi$ and the operator $\odot$ to get the decoded coefficient is termed as fingerprinted coefficient $d_n$.

$$d_n = e_n \odot k_i^\xi \qquad (7)$$

**Step 4**. Excepting the coefficient $e_n$, another coefficients $e_i$ $(0 \le i < n)$ are decrypted to obtain the another decrypted element $d_i$ from bellow formula:

$$d_i = e_i \odot e_{i+1}, i = 0,1,2,\ldots,n\text{-}1 \qquad (8)$$

Here, we also require $(n-1)$ operators $\odot$ that are randomly generated by key $k_i^\Psi = \{-1,1\}$. Operator $\odot$ is also selected the subtraction (-) if $k_i^\Psi = -1$ or the addition (+) if $k_i^\Psi = +1$. Note that, because of $k_i^\Psi \ne k_C^\Psi$, operators $\odot$ patterns that are implemented in the decoding process is difference from patterns using in the encoding process and then it is clear that $d_i \ne p_i$.

Therefore, according to combination of $\{k, k_i\}$, we can implement the incomplete cryptography and then control the quality content $P'_i$ to user $R_i$ based on his/her individual information.

### C. Tracing the source of pirated copies

When a producer wishes to check whether the user is a legal user, he/she can extract the hashing information $H'_i$ from $P'_i$ and compare with the hashing information $H_i$ that is extracted from $P_i$. $P_i$ is the decrypted content by $T$ using the same decode key $\{k, k_i\}$. If the hashing information matches his hashing value, the user is a legal user. Conversely, if the hashing information is different from his hashing value, the user is an illegal user. Furthermore, it can specify to trace the source of pirated copies. The purpose of this proposed method is to inform the users about the existence of individual hashing information which can exactly identify users, and limit the illegal redistribution in advance.

## V. Application DCFF on JPEG algorithm

### A. Experimental environment and image

All experiments were performed by incomplete encoding and incomplete decoding on JPEG [15] images using the Vine Linux 3.2 system. In order to generate the encryption key and decryption key $\{k, k_i\}$, we used function *rand*() of GCC version 3.3.2 (http://gcc.gnu.org/) with seed=1. Additionally, the ImageMagick version 6.6.3-0 (http://www.imagemagick.org/script/) was used to convert and view the experimental JPEG images.

We prepared some different features of experimental images regarding CG, scenery, construction and person. Ten test images are the 8-bit RGB images of SIDBA (Standard Image Data BAse) international standard image (Lighthouse, Pepper, Title, Lenna, Girl, Airplane, Parrots, Couple, Milkdrop, Mandrill) of size $256 \times 256$ pixels.

Here, all images were compressed with quality 75 (the lowest $0 \Leftrightarrow 100$ the highest) to make experimental JPEG images for evaluation of the proposal method.

### B. Evaluation of image quality

We used PSNR (Peak Signal to Noise Ratio) [14] to evaluate the JPEG image quality. In these experiments, the PSNR were calculated with RGB pixel data of original image and the JPEG image. A typical value for PSNR in a JPEG image (quality 75) is about 30dB and the suitable PSNR of trial image is around 20dB [11].

### C. Scrambled content

To make scrambled and incomplete decoding contents of JPEG, and we have selected the quantized DCT tables to implement the incomplete cryptography. In order to simply implement the demonstrated method, we assume that the operators of the encoded key $k_C^\Psi$ are all negative (subtraction: -), and the operators of the decoded key $k_i^\Psi$ are all positive (addition: +) (see **Fig. 3**). The operators of the encoded key $k_C^\Psi$ and the decoded key $k_i^\Psi$ affect the quality of contents in the system.

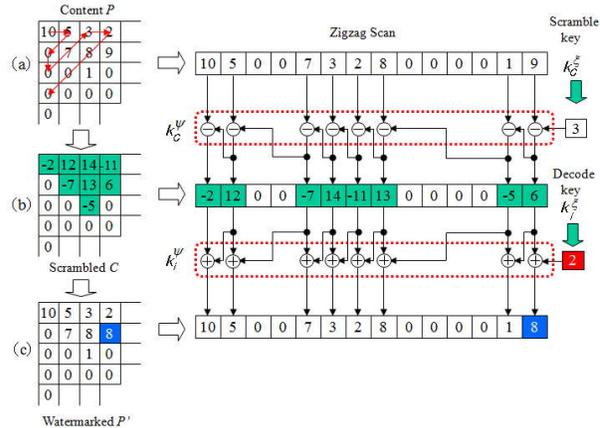

Fig. 3. An example of DCFF algorithm.

First, various parameters such as the side information, entropy code, and so on, are extracted from JPEG data. Suppose that specified DCT table $P$ is selected. An example of implementation method is shown in Fig. 3(a). The zigzag scan result of this DCT table is $\{10,5,0,0,7,3,2,8,0,0,0,0,1,9\}$. To encode the DCT table, $T$ generates an encoded key $k_C^\xi$. In example Fig. 3(a), $k_C^\xi = 3$. After that, the last non-zero coefficient "9" is encoded by replacing with the difference "+6", which is obtained after applying the calculation as "9-3". Excepting the last non-zero coefficient, the non-zero coefficients of $\{10,5,0,0,7,3,2,8,0,0,0,0,1\}$ are encrypted with the previous difference to obtain the scrambled content $C$. If the difference of non-zero coefficient and the previous difference is zero, it is replaced by the closed non-zero coefficient. The scrambled of this example is {-2,12,0,0,-7,14,-11,13,0,0,0,0,-5,6}. This scrambled method is repeated for all DCT tables. The scrambled content $C$ is distributed via network to users for trial.

### D. Fingerprinted content

Owing to the fingerprinting key $k^{\psi_i}$ that is created by user's information, the fingerprinted information is embedded into the decoded content and it is used for tracing the illegal source when fingerprinted content is redistributed via network.

In this decoding process, all encoded non-zero coefficients are not completely decoded. However, the scrambled content $C$ is decoded to close quality of the original content. In the example Fig. 3(b), $C$ is decoded by user $R_i$ depending on the decoded key $k^\xi_i$. Note that, the last non-zero coefficient will be the position that is embedded the fingerprinted information. Assuming that $R_i$ received the key $k^\xi_i = 2$, which is delivered from $T$. Next, $k^\xi_i$ is applied to decode the last non-zero coefficient "6" in the zigzag scan results and the decoded coefficient "8" is acquired (see Fig. 3(b)). Finally, the remaining of non-zero coefficients are decoded with replacing by the addition or subtraction between the non-coefficient and previous difference. As described in Fig. 3(c), the decoded DCT table is {10,5,0,0,7,3,2,8,0,0,0,0,1,**8**}.

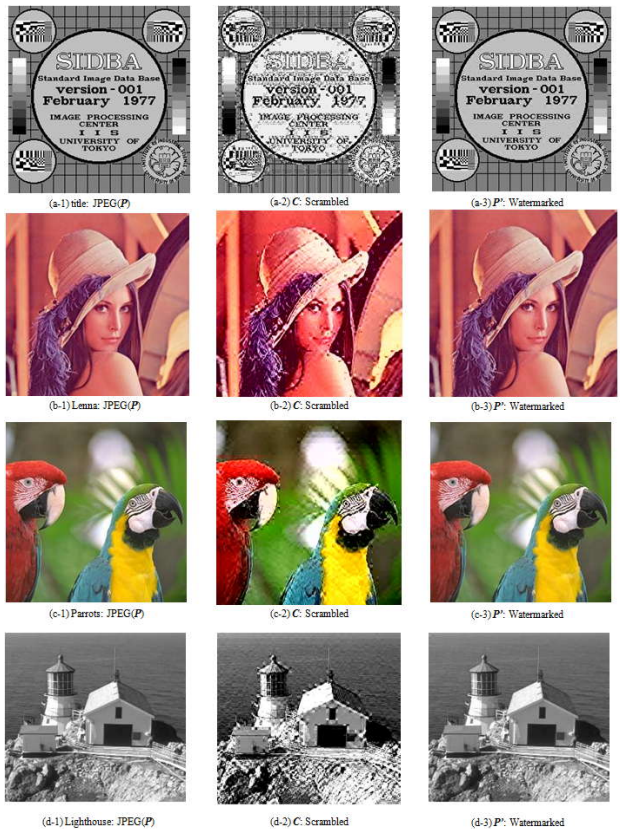

Fig. 4. Examples of experimental images.

The last coefficient **8** in the zigzag scan of Fig. 3(c) is the coefficient in which the userID is embedded.

After obtaining $P'_i$, $P'_i$ is not only decoded with slight deterioration, but also fingerprinted with particular information (i.e. user information). Therefore, if producer $T$ extracts the hash value $H'_i$ (the length of $H'_i$ is 128 bits) from $P'_i$ to exactly identify users.

### E. Experimental results

In the simulated results, firstly, we generated the sets of key $\{k^\xi_C=3, k^\psi_C=$all (-)$\}$, $\{k^\xi_i=2, k^\psi_i=$all (+)$\}$ to encode and decode the original content and the scrambled content. We calculated the PSNR value of the output JPEG images in every processes and extracted the fingerprinting information as hash value (128 bits) from the incomplete decoded JPEG images to identify individual legal user. The experimental results are shown in **Table II**. We see that the fingerprinted JPEG images are not distinguishable from the original JPEG images. The scrambled JPEG images were degraded about 20dB, and they seemed appropriate as trial content.

In addition, **Fig. 4** shows some experimental images. According to the results in Fig. 4, it is possible to create the scrambled content (see Fig. 4 (a-2), (b-2), (c-2), (d-2)) and incomplete decode content (see Fig. 4 (a-3), (b-3), (c-3), (d-3)) based on the incomplete cryptography using DCFF.

TABLE II.    PSNR[dB] AND HASH BITS.

|          | $P$   | $C$   | $P^J$ | $H_i$ |
|----------|-------|-------|-------|-------|
| Airplane | 30.20 | 19.42 | 30.17 | 128   |
| Girl     | 32.71 | 25.38 | 32.68 | 128   |
| Parrots  | 34.25 | 25.57 | 34.24 | 128   |
| Couple   | 34.06 | 27.48 | 34.04 | 128   |
| Title    | 31.84 | 15.17 | 31.82 | 128   |
| Lenna    | 32.37 | 24.37 | 32.36 | 128   |
| Milkdrop | 31.99 | 23.88 | 31.97 | 128   |
| Mandrill | 24.97 | 21.63 | 24.96 | 128   |
| Lighthouse | 32.66 | 23.02 | 32.65 | 128   |
| Pepper   | 28.81 | 22.57 | 28.80 | 128   |

According to the experimental results, we have established the incomplete cryptography system based on the proposed method. The scrambled content is created to disclose the original content and distributed widely to users. In the incomplete decoding process, we changed the some quantized DCT coefficient itself instead of the closed quantized DCT coefficient by an individual decryption key using DCFF. Thus, the original content is not decoded temporarily inside the system. We can conclude that the above technical problem by the conventional DRM system is solved by using the incomplete cryptography system.

### F. Discussion

In this study, we do not focus on the robustness of fingerprinting method. We concentrate to solve the problem of conventional DRM system which is to completely disclose the original content inside user's system while decoding process. We combine two processes (decoding and fingerprinting) at user side to become the incomplete decoding and the user's information is fingerprinted into decoded content simultaneously. Assuming that there are not any attacks on the fingerprinted content, we always identify exactly the legal user by extracting 128 bits hash value $H'_1$ from the fingerprinted content. From this idea, we make the following contributions in this paper:

(1) We proposed the fundamental incomplete cryptography which differs from complete cryptography (e.g. DES, AES, ...). It is promising to be able to solve the problem of conventional DRM system.

(2) We presented a new DRM system that includes trial contents for advertisement and user ID for distinguishing the legal user. Our system makes it easier for users to try the digital content before deciding whether to buy it or not.

(3) Our proposed method can detect the source of pirated content by comparing the extracted hash value from incomplete decoded content with producer's database. It is considered that it can limit the illegal redistribution in advance.

## VI. CONCLUSION

In this paper, we have presented a scheme of an incomplete cryptography system using DCFF method. This approach integrates the encoding process and fingerprinting progress of DRM technology. By doing so, we can eliminate the problem of the present DRM technology and manage the legal user effectively with fingerprinted information individually.

As previously presented in the experiments, it is obvious the above technical problem by the conventional DRM system is solved by using the incomplete cryptography system and proposed method can maintain a good performance in transparency and high quality of decoded content.

The development of the robust incomplete cryptography is our future work.

## ACKNOWLEDGMENT

This work was supported by A.N.Lab JSC.

## REFERENCES

[1] G. H. Kim, D. K. Shin, D. G. Shin, "An Efficient Methodology for Multimedia Digital Rights Management on Mobile Handset," IEEE Trans. on Consumer Electronics, Vol. 50, No. 4, pp. 1130--1134, 2004.

[2] Q. Liu, R. Safavi-Naini, N. P. Sheppard, "Digital Rights Management for Content Distribution," Proc. of AISW2003, Vol. 21, pp. 49-58, 2003.

[3] DRM technology, "Advanced Image Seminar 2003," The Institute of Image Electronics Engineers of Japan, 2003.

[4] V. Wahadaniah, Y. L. Guan, H. C. Chua, "A New Collusion Attack and Its Performance Evaluation," Proc. of IWDW, pp. 88-103, 2002.

[5] D. Kirovski, H. S. Malvar, Y. Yacobi, "Multimedia Content Screening Using a Dual Watermarking and Fingerprinting System," ACM Multimedia, 2002.

[6] W. Trappe, M. Wu, K. J. R. Liu, "Collusion-Resistant Fingerprinting for Multimedia," Proc. Of IEEE Int. Conf. on Acoustics, Speech, and Signal Processing (ICAPSSP'02), Vol. 4, pp. 3309--3312, 2002.

[7] D. Kundur, K. Karthik, "Video fingerprinting and encryption principles for digital rights management," Proc. IEEE vol. 92, no. 6, pp.918--932, 2004.

[8] B.M. Macq, J.J. Quisquater, "Cryptology for digital TV broadcasting," Proc. IEEE vol.83, no. 6, pp.944--957, 1995.

[9] F. Hartung, B. Girod, "Digital watermarking of MPEG-2 coded video in the bitstream domain," Proc. of the IEEE International Conference on Acoustics, Speech and Signal Processing, vol. 4, pp. 2621--2624, 1997.

[10] J. Bloom, "Security and rights management in digital cinema," Proc. of the IEEE International Conference on Acoustics, Speech and Signal Processing, vol. 4, pp. 712--715, 2003.

[11] M. Iwakiri, Ta Minh Thanh, "Fundamental Incomplete Cryptography Method to Digital Rights Management Based on JPEG Lossy Compression," The 26th IEEE International Conf. on Advanced Information Networking and Applications (AINA-2012), 2012.

[12] M. Iwakiri, Ta Minh Thanh, "Incomplete Cryptography Method Using Invariant Huffman Code Length to Digital Rights Management," The 26th IEEE International Conf. on Advanced Information Networking and Applications (AINA-2012), 2012.

[13] J. W. Bos, O. Ozen, and M. Stam, "Efficient hashing using the AES instruction set," In Proc. of the 13th CHES'11, Springer-Verlag, Berlin, Heidelberg, pp.507--522.

[14] K.Matsui, "Fundamentals of Digital Watermarking," Morikita-publisher,1998.(in Japanese)

[15] The International Telegraph and Telephone Consultative Committee Information Technology - Digital Compression and Coding of Continuous-tone still Images - Requirements and Guidelines, International Telecommunication Union, 1992.